\documentclass[twocolumn]{article}
\usepackage{graphicx,amsmath}
\newcommand{\be}{\begin{equation}}
\newcommand{\ee}{\end{equation}}

\newcommand{\A}{{\cal A}}

  \renewenvironment{thebibliography}[1]{%
    \begin{oldthebibliography}{#1}%
      \setlength{\parskip}{0ex}%
      \setlength{\itemsep}{0ex}%
  }%
  {%
    \end{oldthebibliography}%

  }

\begin{document}

\title{Local graph alignment and motif search in
biological networks}

\author{Johannes Berg and Michael L\"assig \\
Institut f\"ur Theoretische Physik,
Universit\"at zu K\"oln, \\
Z\"ulpicherstr. 77, 50937 K\"oln, Germany\\
}
\maketitle

{\it PNAS}, {\bf 101} (41) 14689-14694 (2004),


{\bf \small 
Interaction networks are of central importance in
post-genomic molecular biology, with increasing amounts
of data becoming available by high-throughput methods.
Examples are gene regulatory networks or protein interaction
maps. The main challenge in the analysis of these
data is to read off biological functions from the topology of
the network.
Topological {\em motifs}, i.e., patterns  
occurring repeatedly at different positions in the network
have recently been identified as basic modules of
molecular information processing. In this paper, we discuss
motifs derived from families of mutually similar but not
necessarily identical patterns. We establish a statistical
model for the occurrence of such motifs, from which
we derive a scoring function for their statistical
significance. Based on this scoring function, we develop
a search algorithm for topological motifs
called {\em graph alignment}, a procedure with some analogies
to sequence alignment. The algorithm is applied to the
gene regulation network of {\em E. coli}. }
\vspace{10 pt}


The vast amount of sequence data collected over the past
two decades is at the heart of quantitative molecular
biology. Biological information is extracted from these
data mainly by analyzing similarities between sequences.
This approach is based on efficient {\em sequence alignment}
algorithms and a statistical theory to assess the significance of
the results (see \cite{durbin}).
Its ultimate goal is to infer functional
relationships from correlations between sequences. Over
the last few years, however, it has become clear that
functions in many cases cannot be identified at the level
of single genes. A given function may require the
cooperative action of several genes, and conversely, a
given gene may play a role in quite different functional
contexts. The genome is thus a highly interactive system and
the expression of a gene depends on the activity of other
genes. The pathways of these interactions are encoded in
so-called {\em regulatory networks}. Similarly complex
networks govern {\em signal transduction}, that is, the
influence of external signals on gene expression, or {\em
protein interactions}, that is, the ability of two or more
proteins to form a bound state in a living cell.

A few exemplary cases of gene networks have
been studied in much detail, such as the regulation of
early development in the sea urchin
{\em Strongylocentrotus purpuratus} \cite{davidson} or
in {\em Drosophila}~\cite{tautz.review}.
In some approximation, these structures can be
understood as {\em logical networks}: the expression level
of a gene is reduced to a binary variable ({\em on} or
{\em off}) and is specified in terms of binary input data,
i.e., the expression levels of its ``upstream'' genes.

On the other hand, a large amount of data on molecular
interaction networks is now obtained by high-throughput
experiments, for example protein interaction maps in
yeast~\cite{uetz} or gene expression arrays~\cite{lockhart}.
In these arrays, one probes the activity of an entire
genome, rather than of just a few genes. However, the
detailed logical connection of interaction pathways is
typically lost. The
information is reduced to a {\em topological network}, that
is, a set of nodes (representing, e.g., genes or proteins)
and links representing their pairwise interactions. These
links can be {\em directed} as in the case of regulatory
interactions or {\em undirected} as for protein-protein
binding. The amount of topological data on molecular
networks is expected to increase rapidly in the next few
years, paralleling the earlier explosion of sequence data.

What can be learned from these data? Using the network
topology alone, can we distinguish patterns of biological
function from random background? The purpose of this paper
is to develop a ``bioinformatics'' approach to the search 
for local modules in networks. We discuss a heuristic 
{\em search algorithm}
and its statistical grounding in a stochastic model of 
{\em network evolution}. This approach is designed to complement
experiments in specific organisms by large-scale database searches. 

In two seminal studies~\cite{aloncoli,alonyeast}, it has
recently been shown that topological networks indeed
contain statistically significant patterns indicative of
biological functions. These {\em motifs} are patterns 
which occur more frequently in the observed network than
expected in a suitable null ensemble. The
motifs found so far have been identified because they occur
{\em identically} at different positions in a network.

If network evolution is a stochastic process, however, functionally
related motifs do not need to be topologically identical. Hence, the
notion of a motif has to be generalized to a stochastic one as well.
Variations arise due to uncertainties in the network data, or -- more
importantly -- because some of the interactions can change without
affecting the functionality of the motif. This ``noise'' is an
important characteristic of biological systems, familiar from sequence
analysis where one searches for local sequence similarities blurred by
mutations and insertions/deletions, rather than for identical
subsequences. It leads us to the notion of a {\em probabilistic
  motif} where each link occurs with a certain likelihood.
Probabilistic motifs arise as consensus from finding a family of
``sufficiently'' similar subgraphs in a network. The search for
mutually similar subgraphs and their probabilistic motifs is the
central issue of this paper.

The motifs of interest here
are non-random in two ways: they have an enhanced number
of internal links, associated e.g. with feedback, and they 
appear in a significant number
of subgraphs. Identifying these {\em local} deviations
from randomness in networks requires a 
statistical theory of local graph structure, 
which we establish in this paper. This is a complementary approach
to the global statistics measured by the
connectivity distribution~\cite{newman_p}
or connectivity correlations~\cite{newman_q,berglassig} of a network.

Our approach leads to an algorithmic procedure
termed {\em local graph alignment}, which is conceptually
similar to sequence alignment. It is based on a
{\em scoring function} measuring the statistical
significance for families of mutually similar subgraphs.
This scoring involves quantifying the significance
of the individual subgraphs as well as their mutual
similarity, and is thus considerably more complicated
than for families of identical motifs. Our scoring
function is derived from a stochastic model for network
evolution. There is indeed evidence that network
evolution can be described as a stochastic process. For 
example, the comparison of the regulatory networks for
early development in several {\em Drosophila} species 
has revealed the continuous buildup and loss of gene 
interactions following an approximate molecular clock~\cite{molclock}.
Yet little is known about the specific pathways of network evolution.
Our scoring function is compatible with divergent evolution
of subgraphs but also with convergent evolution towards a common
functional motif. These pathways can be illustrated by a 
comparison with sequence evolution. An example of convergent evolution 
is the formation of sequence motifs serving as  
binding sites of specific enzymes~\cite{stormo,bussemaker}. 
An example of divergent evolution is a set of sequences 
stemming from a common ancestor undergoing mutations independently.
The probabilistic grounding
of graph alignment allows us to infer optimal scoring
parameters by a maximum-likelihood procedure~\cite{felsenstein}. 

As a computational problem, graph alignment is more
challenging than sequence alignment. Sequences can be
aligned in polynomial time using dynamic programming
algorithms. For graph alignment, a polynomial-time algorithm
probably does not exist.
Already simpler graph matching problems such as the
{\em subgraph isomorphism problem} (deciding whether a graph contains a 
given subgraph)~\cite{ullmann,messmerbunke} or finding the {\em
largest common subgraph} of two graphs~\cite{gareyjohnson}
are $NP$-complete and $NP$-hard, respectively. Thus, an
important issue for graph alignment is the construction of
efficient heuristic search algorithms. Here we solve this 
problem by mapping graph alignment onto a spin model familiar 
in statistical physics, which can be treated by simulated annealing. 

This paper is structured as follows. In the first part, we discuss the
statistics of local subgraphs based on a probabilistic model. This is
done in three steps: (i) an individual subgraph with an enhanced
number of internal links, (ii) a subgraph in the presence of a
template motif specifying the functional importance of each link, and
(iii) correlated subgraphs, whose common pattern is to be inferred
from the data instead of being given as a template.  We then construct
a scoring function designed to distinguish sets of statistically
significant network motifs with an enhanced number of links from a
background of other patterns.  High-scoring motifs are found by an
alignment algorithm, details of which are described in the 
supporting text.  In the second part of the paper, we
apply this method to the regulatory network of \emph{Escherichia coli}
and discuss the probabilistic motifs found. The statistics of these
motifs is used to test the assumptions of our probabilistic model.

\subsubsection*{Graphs and patterns}

A topological network or {\em graph} is a set of {\em
nodes} and  {\em links}.  Labeling the nodes by an index $r
= 1, \dots, N$, the network is  described by the {\em
adjacency matrix} ${\bf C}$, which has entries  $C_{r r'}
= 1$ if there is a directed link from node $r$ to node $r'$  and
$C_{rr'} = 0$ otherwise. Graphs with a generic
adjacency matrix are called {\em directed}. The
special case of a symmetric adjacency matrix can be used to
describe {\em undirected} graphs. The {\em in} and  {\em out
connectivities} of a node, $k^{+}_{r} = \sum_{r'}
C_{r'r}$  and $k^{-}_{r} = \sum_{r'} C_{r r'}$, are
defined as the number of in- and outgoing links,
respectively. The total number of links is denoted by
$K = \sum_{r,r'} C_{rr'}$. The networks considered
here are {\em sparse}, i.e., their average
connectivity $K/N$ is of order $1$.

A {\em subgraph} $G$ is given by a subset of $n$ vertices $\{r_1,
\dots r_n \}$ and the resulting restriction of the adjacency matrix.
More precisely, we define the matrix ${\bf c}(G, \A)$ with the entries
$c_{ij} = C_{r_i r_j}$ ($i,j = 1, \dots, n$) specifying the internal
links of the subgraph for a given order $\A$ of the nodes. This matrix
${\bf c}$ is called a {\em pattern}, which is contained in the
subgraph.  The definition of a pattern used here implies that two
patterns are counted as separate if the matrices ${\bf c}$ and ${\bf
  c}'$ are different. This assumes that nodes are distinguishable by
their biochemical identity and their functional role even if they are
at symmetric positions, i.e., if ${\bf c}$ and ${\bf c'}$ differ only
by the labeling of the nodes. An alternative definition would count
two matrices ${\bf c}$ and ${\bf c'}$ related by a relabeling as
defining an identical pattern. Which definition is more appropriate
depends on the particular biological application.
  
  The most important characteristic
of patterns for what follows is their number of internal links, \be L
({\bf c}) = \sum_{i,j} c_{ij}.
\label{L}
\ee
Fig.~\ref{figmotifs} shows two subgraphs that differ
in the values of $L$. 

\subsubsection*{Graph alignments and motifs}

\begin{figure}[th!]
\includegraphics*[width=1 \linewidth]{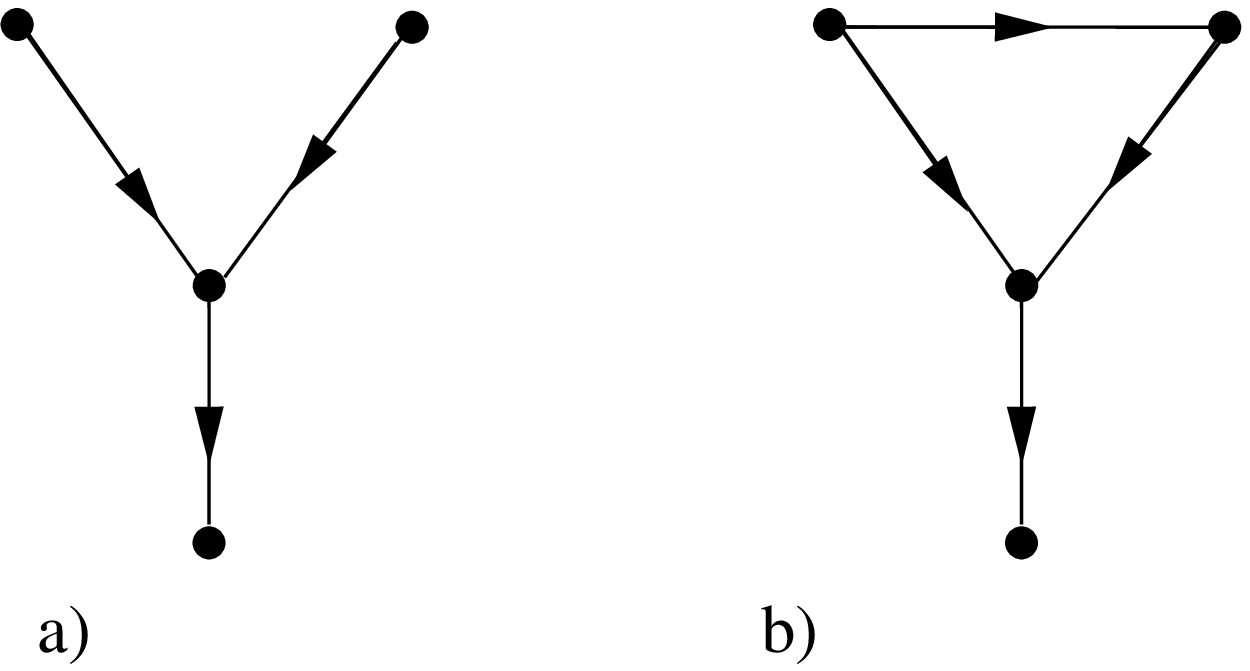}
\vspace*{.5cm}
\includegraphics*[width=1 \linewidth]{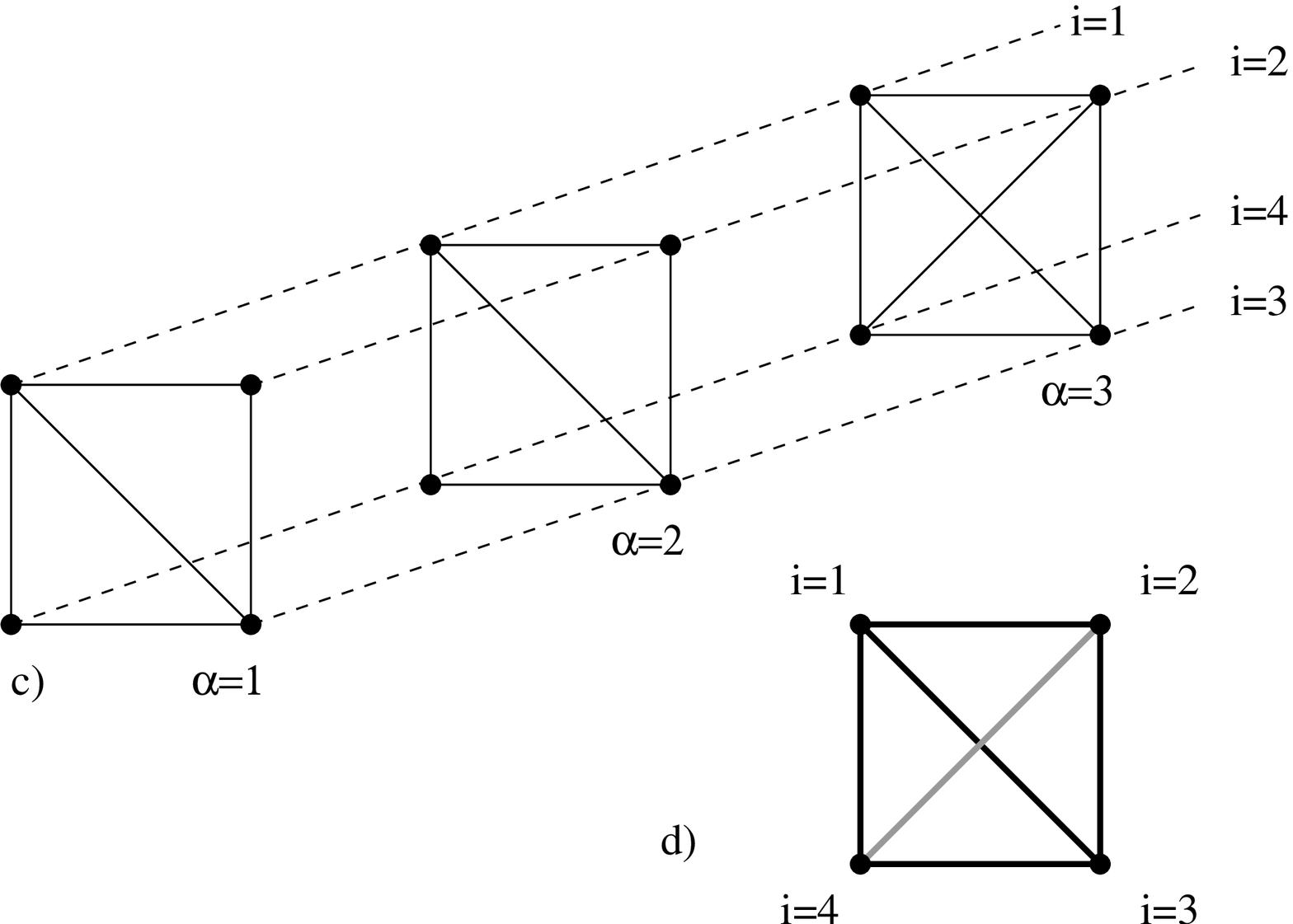}
\caption{\label{figmotifs} \small
{\bf Motifs and alignment in topological networks.}
(a)~A randomly chosen connected subgraph is likely to be a tree, 
i.e., it has a the number of internal links links equal to its number 
of nodes minus 1. 
(b)~Putatively functional subgraphs are distinguished by internal loops,
i.e., by a higher number of internal links. 
(c)~An alignment of three subgraphs with four nodes each. 
Each nodes carries an index $\alpha =1,2,3$ labeling its subgraph
and an index $i = 1,2,3,4$ given by the order of nodes within the 
subgraph. Nodes with the same index $i$ are joined by dashed lines,
defining a one-to-one mapping between any two subgraphs. Network
links are shown as solid lines (with their arrows suppressed for 
clarity). 
(d)~The consensus pattern of this alignment. Each link
occurs with a likelihood $\bar{c}_{ij}$ indicated by the gray scale.
}
\end{figure}

A {\em graph alignment} is defined by a set of several
subgraphs  $G^\alpha$ ($\alpha = 1, \dots, p$) and a
specific order of the nodes $\{r_1^\alpha, \dots, r_n^\alpha\}$
in each subgraph; this joint order is again denoted by $\A$. For
simplicity, we assume here that the subgraphs are of the same size $n$,
but it is not difficult to generalize our approach in order to include
subgraphs of different size.  For a given set of $p$ 
mutually disjoint subgraphs, there are 
$(n!)^p$ different alignments.
An alignment associates each node in a subgraph with exactly
one node in each of the other subgraphs. This can be visualized by $n$
``strings'', each connecting the $p$ nodes with the same
index $i$ as shown in fig.~\ref{figmotifs}~(c).

A given alignment $\A$ 
specifies a pattern in each subgraph, we write
${\bf c}^\alpha \equiv {\bf c} (G^\alpha, \A)$. 
The {\em consensus pattern} of this  alignment is given
by the matrix
\be
\label{cbar}
\overline {\bf c} = \frac{1}{p} \sum_{\alpha =
1}^p {\bf c}^\alpha.
\ee
This is a {\em probabilistic pattern}, the entry $\overline c_{ij}$
denoting the likelihood that a given link is present in the aligned
subgraphs.
For any two aligned subgraphs $G^\alpha$ and $G^\beta$, we can define
the {\em pairwise mismatch}
\begin{equation}
\label{pairwise_mismatch}
M({\bf c}^\alpha, {\bf c}^\beta) =
  \sum_{i,j=1}^n [c_{ij}^\alpha (1 - c_{ij}^\beta) +
            (1 - c_{ij}^\alpha) c_{ij}^\beta ] \ .
\end{equation}
The mismatch is $0$ if and only if the matrices ${\bf
c}^\alpha$ and ${\bf c}^\beta$ are equal, and is positive
otherwise.  It can be considered as a Hamming distance 
for aligned subgraphs. The average mismatch
over all pairs of aligned subgraphs,
$\overline M \equiv M(\overline {\bf c}, \overline {\bf c})$, 
is termed the {\em fuzziness} of the consensus
pattern $\overline {\bf c}$. Analogously, the 
average number of internal links is 
denoted by $\overline L \equiv L(\overline {\bf c})$.

We now define {\em network motifs} as statistically significant 
consensus patterns of graph alignments, which are distinguished 
by a high number of internal links and low fuzziness. Clearly,
this definition is mathematically loose before we quantify 
the statistical significance. This will be done in the next three 
sections. 

Guided by the results of refs.~\cite{aloncoli,alonyeast},
we take an enhanced number of internal links as a
topological indicator of possible functional modules in
networks. The additional links beyond a treelike topology 
can be associated with
feedback or feed-forward loops in transcription networks, or clusters in
protein interaction networks. 
For example, the triangle shown in 
fig.~\ref{figmotifs}~(b) can be interpreted~\cite{aloncoli} 
as a low-frequency bandpass
filter: the central node is activated if both top nodes are active.
However, the right hand node is activated by that on the left 
with a small delay,
so the central node is activated provided the left node is active
for a time {\em longer} than this delay. The non-treelike nature of 
this motif is crucial for its function. 
On the other hand, most randomly
chosen {\em connected} subgraphs would be treelike. 
Clearly, an enhanced number of internal links is but the simplest
topological indicator of putative functionality, and more detailed
ways of identifying network motifs are likely to emerge in the future. 

\subsubsection*{Statistics of individual subgraphs}

In order to quantify the statistical significance of
a given number of internal links, we first compute the
relevant probability distribution in a suitable random graph ensemble,
which is generated by an unbiased sum over all graphs with the
same number of nodes and the same connectivities $k^{-}_r,k^{+}_r$ 
($r=1,\ldots,N$) 
as in the data set but randomly chosen links~\cite{maslovsneppen,berglassig}.
This {\em null ensemble} is appropriate for biological networks, whose 
connectivity distribution generally differs markedly from that 
of a random graph with uniformly distributed links. 
In the null ensemble, the probability of finding a
directed link from node $r_i$ to node $r_j$ is in good approximation given by
$w_{ij}=k^{-}_{r_i} k^+_{r_j}/K$~\cite{alonsubgraph}. Hence,
a given subset of nodes $\{r_1, \dots, r_n \}$ forms a
subgraph $G$ with probability 
\be
\label{single_null}
P_0 (G) = \prod_{i,j=1}^n
                (1-w_{ij})^{1-c_{ij}}  w_{ij}^{c_{ij}}.
\ee
This expression neglects double links,
which can be included as in~\cite{alonsubgraph}. 
The probability $P_0(G)$ depends on the pattern
${\bf c}(G)$ of the subgraph, 
as well as on its environment given by the connectivities
$k^+_i, k^-_i$ ($i = 1, \dots, n$). In this ensemble,
the expected number of internal links per node is small,
$\langle L \rangle_0 /n \sim  n / N$, where we denote the 
average over a given ensemble by $\langle\rangle$. Hence, most random
subgraphs in a large and sparse graph are disconnected.
Within the subset of connected subgraphs, most are treelike. (Later 
we will be interested in the subset of non-treelike subgraphs, and 
this will require a modification of the null ensemble.) 

We now assume that subgraphs containing network motifs
are generated by a different ensemble $P_{\sigma}(G)$. The probability 
that a given
pair of nodes carries a link is enhanced by a factor $\mbox{e}^{\sigma}$
relative to the null ensemble (\ref{single_null}),
leading to 
\be
\label{sigma_enhanced}
P_{\sigma} (G) / P_0 (G)= Z_{\sigma}^{-1} \exp [\sigma L({\bf c})]\ .
\ee
Again the probability $P_{\sigma}(G)$ that a given subset of nodes 
$\{r_1, \dots, r_n \}$ forms a subgraph $G$ depends on the 
matrix ${\bf c}(G)$. We have introduced the normalization factor
$Z_{\sigma} = \prod_{ij}\sum_{c_{ij}=0,1} \exp[\sigma L({\bf c})] \, P_0 (G)$, 
which ensures that $P_{\sigma}(G)$ summed over all matrices ${\bf c}$ gives 
unity. The quantity 
$\sigma$, called the {\em link reward}, is multiplied by
the total number $L$ of internal links given by (\ref{L}).
The ensemble (\ref{sigma_enhanced})  is 
a statistically unbiased way to describe 
that functional motifs are distinguished by a large number of internal
links. (Technically, it is the ensemble of maximal information
entropy with a given average link number  $\langle L \rangle$, 
which is determined by the 
value of $\sigma$.) This ensemble may be thought of as resulting from an
evolutionary process favoring the formation of links due
to selection pressure; such a process has recently been
studied for regulatory networks~\cite{radicetal}.
Here we focus on the detection of evolved motifs
rather than on the reconstruction of evolutionary histories.
Hence, we do not need to make assumptions on dynamical
details of motif formation but only on its outcome, which
is described by the ensemble $P_{\sigma}(G)$. We have tested the
form of this ensemble for the regulatory network of {\em
E.~coli} as discussed in the results section
below. Moreover, the value of the link reward $\sigma$
can be inferred from the data. One finds
$\mbox{e}^{\sigma} \sim N/n$, which results in a finite
expected number $\langle L \rangle/n$ of internal links per node
within a motif.

\subsubsection*{Statistics in the presence of a template}

The distribution (\ref{sigma_enhanced}) describes an 
ensemble with an enhanced
number of links, which is appropriate for scoring
individual subgraphs in the absence of further knowledge.
Consider now an evolutionary process directed towards
a given network motif represented by a {\em template}
adjacency matrix ${\bf t}$. An alignment $\A$ between the 
motif ${\bf t}$ and the subgraph $G$ is specified by a
given ordering of the nodes $\{r_1, \dots, r_n \}$ in
$G$. The outcome of this evolutionary process
can be modeled by an ensemble $Q_{\bf t}(G,\A)$ with a bias against links that
do not occur in the template,
\be
\label{template_enhanced}
Q_{\bf t}(G,\A)/ P_0 (G)= Z_t^{-1}
\exp \left [ \sigma L({\bf c}) - \frac{\mu}{2} M({\bf c}, {\bf
t}) \right ] .
\ee
This denotes the probability that a given subset of nodes 
$\{r_1, \dots, r_n \}$ forms an aligned  subgraph $(G, \A)$, 
with the definition (\ref{pairwise_mismatch}) of the pairwise mismatch 
of the subgraph $G$ and the template ${\bf t}$ in a given alignment $\A$. 
Again $Z_t$ is given by normalization. 
This is a {\em hidden Markov model}: the outcome of the stochastic 
process is an aligned subgraph $(G, \A)$ while only $G$ is observed.
The likelihood of observing 
$G$ is then a sum over all alignments,
\begin{equation}
 Q_{\bf t}(G) = \sum_\A Q_{\bf t}(G, \A). 
 \end{equation}
This ensemble has two free parameters, the link reward
$\sigma$ and the mismatch penalty $\mu$ (with a factor
$1/2$ introduced for later convenience). It is conceptionally
similar to hidden Markov models for the alignment of sequences 
with gaps.

\subsubsection*{Statistics of correlated subgraphs}

Now we turn to the case 
where a network motif is not given as a template
but has to be inferred from a family of suitably
aligned subgraphs. The underlying evolutionary
process can be regarded as a biased link formation as in the previous
section, with the consensus pattern $\overline c$ as
``template''. Assuming the link formation is independent
for each subgraph, we obtain an  ensemble given by
\begin{eqnarray}
\label{q_alignment}
\lefteqn{Q_{\sigma,\mu} (G^1, \dots, G^p, \A) / 
\prod_{\alpha = 1}^p P_0 (G^{\alpha})}  \\
& & = Z_{\sigma, \mu}^{-1}
\exp \left [ \sigma \sum_{\alpha = 1}^p L({\bf c^\alpha})
- \frac{\mu}{2} \sum_{\alpha =1}^p M({\bf c}^{\alpha}, {\overline {\bf c}})
\right ] \nonumber
 \\
& & =
 Z_{\sigma, \mu}^{-1}
\exp \left [  \sigma \sum_{\alpha=1}^p L( {\bf c}^{\alpha} )
- \frac{\mu}{2p} \sum_{\alpha,\beta=1}^p
M({\bf c}^{\alpha},{\bf c}^{\beta})
\right ], \nonumber
\end{eqnarray}
where $\A$ specifies an alignment of all subgraphs and 
we have used the definition (\ref{cbar}) of the
consensus pattern. The normalization is given by
\begin{eqnarray}
\label{Z_musigma}
Z_{\sigma, \mu} &=& \sum_{\A} 
\sum_{{\bf c}^1, \dots, {\bf c}^p }
\exp \left [  \sigma \sum_{\alpha=1}^p L( {\bf c}^{\alpha} ) \right.
\\
&  & \left. - \frac{\mu}{2p} \sum_{\alpha,\beta=1}^p
M({\bf c}^{\alpha},{\bf c}^{\beta})
\right ]  
\, \prod_{\alpha = 1}^p
P_0 (G^{\alpha}) . \nonumber 
\end{eqnarray}

\subsubsection*{The scoring function}

We now construct a {\em scoring function} designed to select a set of
(putatively) functional subgraphs -- characterized by a consensus 
motif with a high number of internal links and low fuzziness -- 
from the background of random subgraphs in a large network. 

Based on the preceding discussion, we assume that the statistics of
functional motifs is described by an ensemble $Q(G^1, \dots, G^p,
\A)=Q_{\sigma,\mu}(G^1, \dots, G^p, \A)$, where the scoring parameters
$\sigma$ and $\mu$ remain to be determined from the data. 

For the biological applications described above, where internal 
links are associated with feedback loops, it is clearly useful
to restrict the motif search to the set of all connected subgraphs 
which contain internal loops, i.e., which are non-treelike. 
For connected subgraphs of size $n$, this set is given by the 
constraint $L \geq n$ on the internal link number. 
A large random graph typically contains a number of order
one of such subgraphs, and these define the relevant null ensemble for
motif search. We model these subgraphs using the ensemble $P_{\sigma_0}$ 
with an enhanced number of links defined in (\ref{sigma_enhanced}). 
The parameter $\sigma_0$ will be adjusted such that the average number 
of internal links in the null ensemble equals that found in the 
non-treelike subgraphs of a suitable randomized graph. 
Comparing with the ensemble $P_0$ of random subgraphs introduced
earlier,  it is clear that the constraint $L \geq n$
corresponds to a link reward $\sigma_0 > 0$.

Given these two ensembles, we define the log-likelihood score 
\begin{eqnarray}
\label{align_score}
\lefteqn{S(G^1, \dots, G^p, \A)}
\nonumber \\
&  = & 
\log  \left (\frac{Q_{\sigma,\mu} (G^1, \dots, G^p, \A)}{
            P_{\sigma_0} (G^1, \dots, G^p, \A)} \right)
\nonumber \\        
& = &  (\sigma - \sigma_0) \sum_{\alpha=1}^p L( {\bf c}^{\alpha})
- \frac{\mu}{2p} \sum_{\alpha,\beta=1}^p
M({\bf c}^{\alpha},{\bf c}^{\beta})
 \nonumber \\
 & & -\log (Z_{\sigma,\mu}/Z_{\sigma_0}) \ ,
\end{eqnarray}
which is positive if a set of subgraphs $G^1, \dots, G^p$ and 
an alignment $\A$ between them is more likely to occur in the
ensemble $Q_{\sigma,\mu}$ than in the null ensemble
$P_{\sigma_0}$. The term
$\log(Z_{\sigma,\mu}/Z_{\sigma_0})$ acts as a threshold assigning a
negative score to alignments with too large fuzziness or a too small 
number of internal links.

As is clear from the form of the scoring function, graph alignment is
a nontrivial optimization problem, the statistical weight of each
subgraph $G^\alpha$ depending on the scoring parameters as well as on
the other subgraphs included in the alignment. We address this problem
in two steps. First we find the maximum-score alignment(s) for given
score parameters, which is essentially an algorithmic search problem.
Then we discuss the parameter dependence of high-scoring alignments
and obtain the optimal values of $\sigma$ and $\mu$ for a given data
set from a maximum-likelihood procedure.

\subsubsection*{Maximum score alignments and parametric
optimization}

Finding the maximum score alignments
involves a huge search space of possible alignments. The number of alignments 
is of order $(np)^N$ for given $p$ and the computational expense grows 
further when the optimization over $p$ is performed.
Here we use a heuristic algorithm, which can
be described by a mapping to a discrete spin model. First we
enumerate all non-treelike subgraphs of $n$ nodes, which
is feasible for modest values of $n$, and label them
by the index $\alpha = 1, \dots, p_{\max}$. Next
we evaluate the internal link numbers $L^{\alpha}=L({\bf c}^{\alpha})$ and 
the pairwise mismatches $M^{\alpha \beta}$,
defined as the minimum of $M({\bf c}^\alpha,{\bf c}^\beta)$ 
over all {\em pairwise} alignments of the
subgraphs $G^\alpha$ and $G^\beta$. High-scoring {\em multiple} alignments
are then found by a simulated annealing algorithm
in the space $(s^1, \dots, s^{p_{\max}})$, where each ``spin''
$s^\alpha$ takes the value $1$ if $G^\alpha$ is included 
in the alignment and $0$ otherwise. The resulting Hamiltonian ${\cal H}$ 
is  
\begin{eqnarray}
\label{hamiltonian}
\lefteqn{-{\cal H}=
(\sigma-\sigma_0) \sum_{\alpha=1}^{p_{\max}} L^{\alpha} s^{\alpha} }\\
&&-\frac{\mu}{2 p} 
\sum_{\alpha,\beta=1}^{p_{\max}} \tilde{M}^{\alpha \beta} s^{\alpha} 
s^{\beta}-\log (Z_{\sigma,\mu}/Z_{\sigma_0}) \ , \nonumber
\end{eqnarray}
where $p=\sum_{\alpha} s^{\alpha}$.  The coupling between $s^{\alpha}$
and $s^{\beta}$ is given by $\tilde{M}^{\alpha \beta}$, which is equal
to the pairwise mismatch $M^{\alpha \beta}$ if subgraphs $\alpha$ and
$\beta$ do not overlap, and a large positive constant if they do. (Two
subgraphs overlap if they have more than one node in common. According
to this definition, links in non-overlapping subgraphs form
independently as assumed in (\ref{q_alignment}).)  The threshold term
$\log (Z_{\sigma,\mu}/Z_{\sigma_0})$ is evaluated by saddle-point 
integration, details are given in the supporting text. 
Simulated annealing using the Hamiltonian (\ref{hamiltonian}) will 
then yield high-scoring 
alignments of non-overlapping subgraphs~\cite{domany}. 

For fixed values of the scoring parameters, 
the algorithm is expected to produce
well-defined maximum-score alignments. This can be understood as
follows.  For a (hypothetical) alignment of subgraphs with equal
number of internal links and equal pairwise mismatches, the score
(\ref{align_score}) scales linearly with $p$, the number of aligned
subgraphs. This is consistent with the interpretation of
(\ref{align_score}) as a log-likelihood score, since the aligned
subgraphs occur independently. A high-scoring alignment in a realistic
network may consist of a limited number of identical or very similar
motifs. As we extend this alignment to include more subgraphs,
subgraphs with increasing mutual mismatches are included.  Hence, we
expect the total mismatch to increase faster than linearly with $p$,
leading to a maximum $S^*(\sigma, \mu)$ of the total score at some
intermediate value of $p^*(\sigma, \mu)$. 

The properties of the maximum-score alignments depend strongly
on the parameters $\sigma$ and $\mu$. With increasing
$\sigma$, the number of internal links $L^* (\sigma, \mu)$ per subgraph
is expected to increase. With increasing $\mu$, both the number of
graphs $p^*(\sigma, \mu)$ and the fuzziness $\overline M^*(\sigma,\mu)$
decrease. In this way, the maximum-score alignment varies between
a set of independent subgraphs for $\mu = 0$ and a set of identical
subgraphs with identical motifs for $\mu \to \infty$.

A maximum-likelihood approach can be used to infer the optimal scoring
parameters $\sigma^*, \mu^*$ for a given data set, which we obtain
as the point of the global score maximum
$S^* = \max_{\sigma, \mu} S^*(\sigma, \mu)$.

\subsubsection*{Results and Discussion}

In this section, we discuss the application of local graph alignment
to motif search in the gene regulatory network of 
{\em E. coli}, taken from 
http://www.weizmann.ac.il/mcb/UriAlon/ \\
Network\_motifs\_in\_coli/ColiNet-1.1/,
containing $424$ nodes and $577$ directed 
links. Each labeled node in this network represents a gene. A directed link 
between two nodes signifies that the product of the gene represented 
by the first node acts as a transcription factor on the gene represented 
by the second. Throughout we consider motifs with a fixed number of 
nodes $n=5$. 

\begin{figure}[t]
(a)
\includegraphics*[width=.8 \linewidth]{pscan.eps}

(b)
\includegraphics*[width=.8 \linewidth]{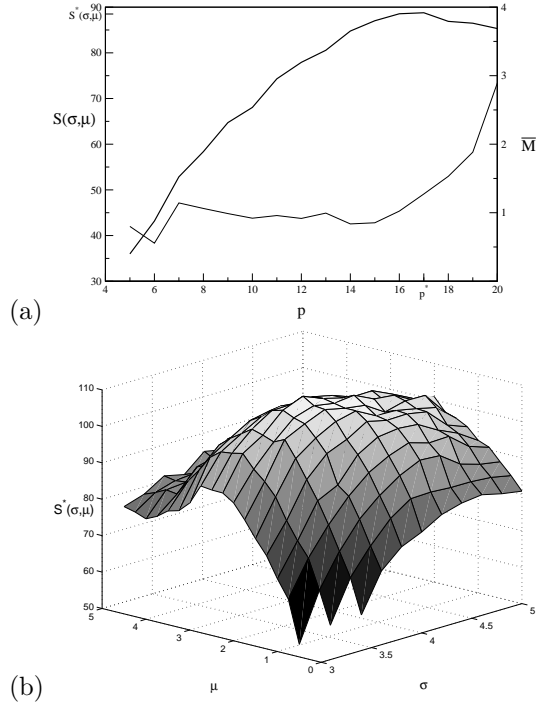}
\caption{ \label{figopt} \small
{\bf Maximum score alignment and parametric optimization}.
(a) Score optimization at fixed scoring parameters $\sigma = 3.8$ and 
$\mu = 4.0$.
The total score $S$ (thick line) and the fuzziness ${\overline M}$
(thin line) are shown
for the highest-scoring alignment of $p$ subgraphs, plotted as a
function of $p$. 
(b)
The score $S^*(\sigma,\mu)$ plotted against the
parameters $\mu$ and $\sigma$. The unique maximum $S^*$ defines the 
maximum-likelihood parameters $\sigma^*=3.8$ and $\mu^*=2.25$. 
}
\end{figure}

First we show that our algorithm indeed produces well-defined
alignments of maximal score (i.e., of maximal relative likelihood).
For fixed parameters $\sigma = 3.8$ and $\mu = 4.0$, this is
illustrated by Fig.~\ref{figopt}~(a), which shows the score $S$ and
the fuzziness ${\overline M}$ for the highest-scoring alignment with a
prescribed number $p$ of subgraphs, plotted against $p$. As expected,
the fuzziness increases with increasing $p$, and the total score
reaches its global maximum $S^*(\sigma, \mu)$ at an intermediate value
$p^*(\sigma, \mu)$. It is lower for $p < p^* (\sigma, \mu)$ since the
alignment contains less subgraphs and for $p > p^* (\sigma, \mu)$
since the subgraphs have higher mutual mismatches.
Fig.~\ref{figopt}~(b) shows the score $S^* (\sigma, \mu)$ as a
function of $\sigma$ and $\mu$. This function has a unique global
maximum $S^*$, which defines the maximum-likelihood point
$(\sigma^*=3.8, \mu^* = 2.25,p^*=24)$.

The scoring parameter $\sigma_0$ of the null ensemble 
$P_{\sigma_0}$ is determined as follows: 
the data set is randomized by generating a network
with the same connectivities as in the data set but randomly chosen
links~\cite{maslovsneppen,berglassig}. Again, the non-treelike subgraphs are 
extracted and their average number of 
internal links is determined. The value of $\sigma_0$ is uniquely 
determined by the condition that the expected number of links in 
the null ensemble (\ref{sigma_enhanced}) equals the average number 
of internal links found in the non-treelike subgraphs. One 
obtains $\sigma_0=2.45$. As expected, $\sigma_0<\sigma^*$, which shows that 
the data set has an enhanced number of internal links relative to 
the randomized network. 

\begin{figure}[t!]
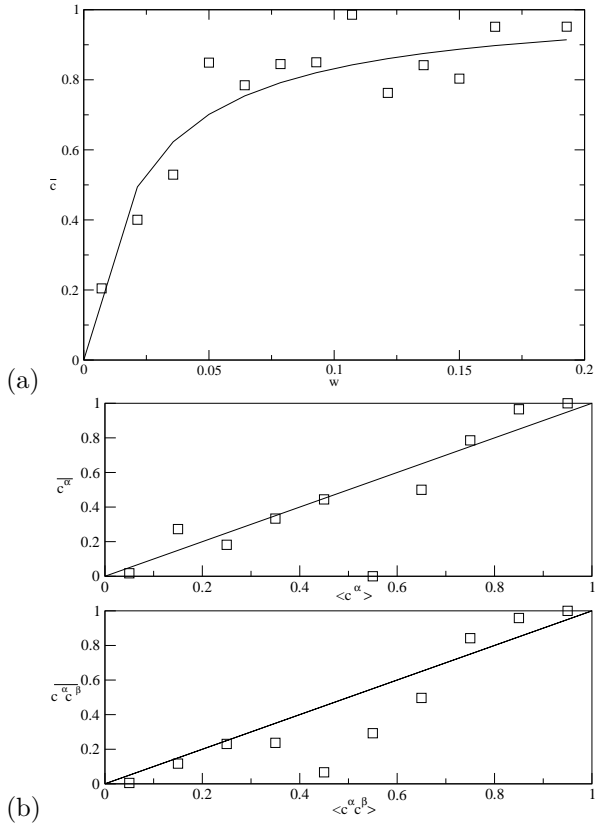

(a)
\includegraphics*[width=.9 \linewidth]{cbar_vs_p.n5.eps}

(b)
\includegraphics*[width=.9 \linewidth]{testmod.eps}
\caption{\label{figtest_mod} \small 
{\bf Statistics of motif ensembles.}
(a) Testing the statistical model for single subgraphs (\ref{sigma_enhanced}).
Non-treelike subgraph are enumerated and node pairs $i,j$ binned
according to $w_{ij}$. The fraction of such pairs carrying a link
is shown against $w_{ij}$. The solid line results from fitting the
model with enhanced number of links
(\ref{sigma_enhanced}) to this data, giving $\sigma=3.8$. 
(b)~Testing the statistical model for alignments (\ref{q_alignment}).
Top: The average value of $c^{\alpha}_{ij}$ over all $\alpha,i,j$ with a
given expectation value of $c^{\alpha}_{ij}$ according to
(\ref{q_alignment}) at
$\sigma=\sigma^*=3.8$ and $\mu=\mu^*=2.25$ against the corresponding 
expectation value (squares). For a perfect fit between model 
and data a straight line
is expected (shown solid).
Bottom: The same procedure is used averaging the two-point function
$c^{\alpha}_{ij}c^{\beta}_{ij}$ over all $\alpha,\beta,i,j$ with a
given expectation value 
$\langle c^{\alpha}_{ij}c^{\beta}_{ij} \rangle$.
}
\end{figure}

At the maximum-likelihood scoring parameters, we can moreover verify 
the functional form of the ensembles used to construct the score function 
(\ref{align_score}). In order to test the
model~(\ref{sigma_enhanced}) for individual subgraphs, we enumerate all
subgraphs with $n=5$ that have non-treelike patterns 
(i.e., a link number $L \geq 5$).
All ordered pairs of nodes $i,j$ are then binned according to the
the probability $w_{ij}$ of a directed link existing between them in the
ensemble $P_0(G)$. In Fig.~\ref{figtest_mod}~(a) the fraction of these pairs
$i,j$ carrying a link is plotted against $w$ (square symbols).
The expectation value of this fraction is given by (\ref{sigma_enhanced})
as ${\rm e}^\sigma w / (1 -w + {\rm e}^\sigma w)$, shown as a solid
line with a fit value $\sigma^* = 3.8$.

Our model (\ref{q_alignment}) for generic alignments can be tested
in a similar way. From this ensemble, the marginal probability that a 
given ordered
pair of nodes specified by $\alpha,i,j$ is linked can be computed.
We group all such pairs with the same
expectation value $\langle c^{\alpha}_{ij} \rangle_{\sigma^*, \mu^*}$
according to (\ref{q_alignment}) to  build a histogram.
For each group, the average of ${c^{\alpha}_{ij} }$ over node pairs in the
actual maximum-likelihood alignment is computed
and plotted against the model prediction, see fig.~\ref{figtest_mod}(b).
The same procedure is repeated for the two-point correlations
$\langle c^{\alpha}_{ij} c^{\beta}_{ij} \rangle_{\sigma^*, \mu^*}$
between associated nodes in different subgraphs $\alpha$ and $\beta$
as also shown in fig.~\ref{figtest_mod}(b). In both histograms, the data points
cluster well around the straight line equating expectation values 
in the model (\ref{q_alignment}) and averages
in the actual alignment. The fluctuations seen reflect the limited
size of the data set and the small number of fitting parameters
in the model. For such data, more detailed models can hardly be
tested since they would lead to overfitting.

\begin{figure}[t!]
(a)
\includegraphics*[width=.96\linewidth]{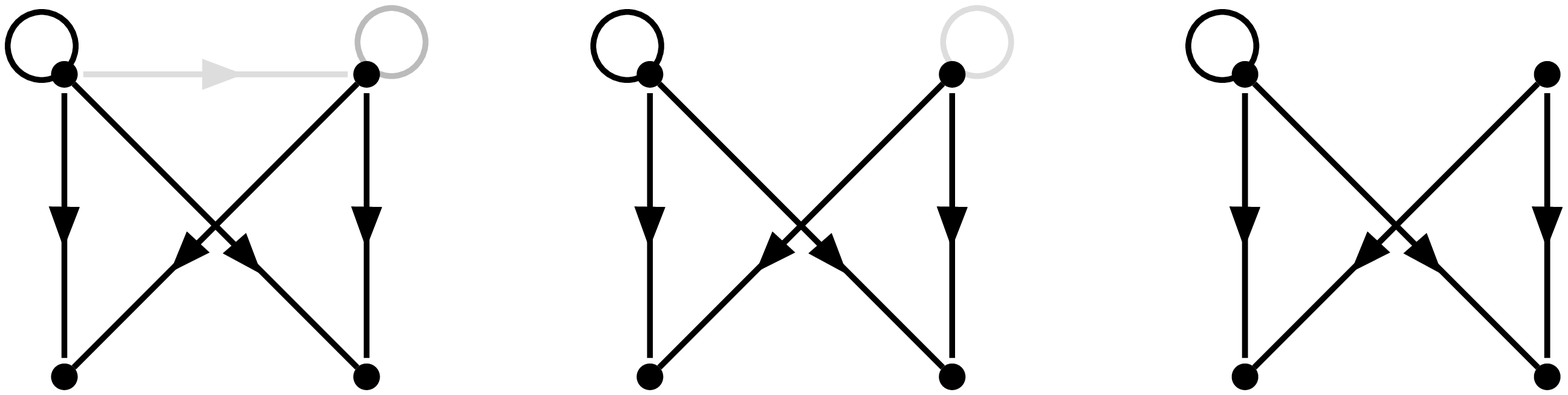}
\vskip0.5cm
(b)
\includegraphics*[width=.96\linewidth]{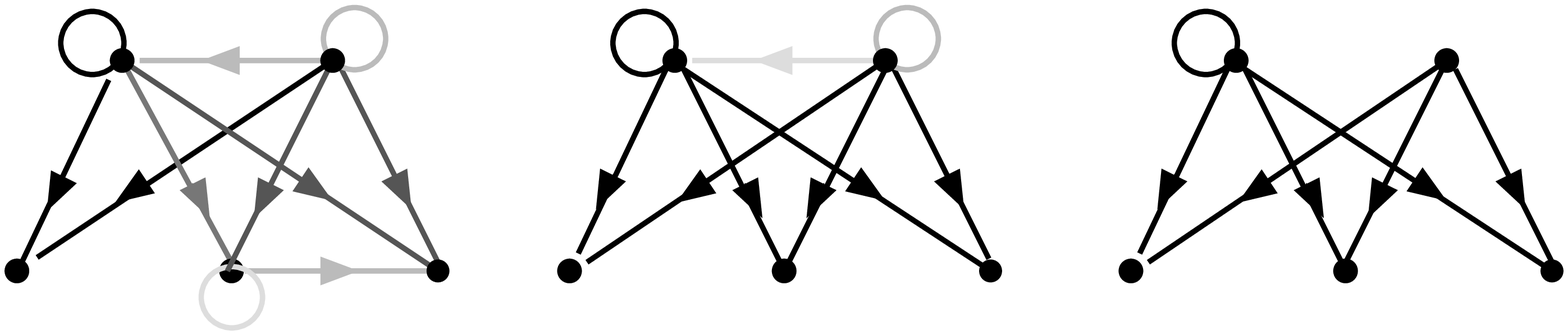}
\vskip0.5cm
(c)
\includegraphics*[width=.5\linewidth]{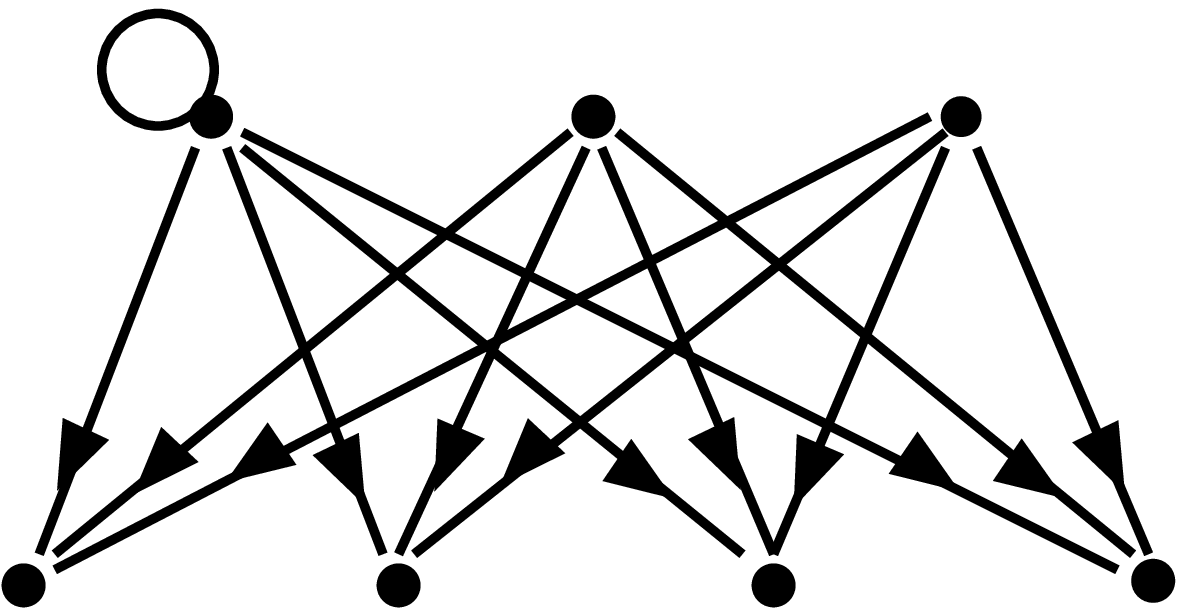}

\vspace{.5cm}
(d)
\includegraphics*[width=.96\linewidth]{muscan.eps}
\caption{\label{figmuscan} \small
{\bf Probabilistic motifs in the E. coli transcription network.}
(a) Consensus motifs with $n=4$ nodes at different values of 
$\mu$. From left to right, $\mu=\mu^*=3.6$, $\mu=8$, and  $\mu=15$. 
The gray-scale of the links indicates the
likelihood that a given link is present in the aligned
subgraphs; the $5$ gray values correspond to $\bar{c}$ in the
range $0.1-0.2$, $0.2-0.4$, $0.4-0.6$, $0.6-0.8$, $0.8-0.9$,
links with $\bar{c}>0.9$ are shown black. 
The link reward is kept fixed at $\sigma=\sigma^*=3.6$ and 
$\sigma_0$ takes on the value $3.15$. 
(b) Consensus motifs with $n=5$ for different 
$\mu=\mu^*=2.25$, $\mu=5,$ and $\mu=12$ (left to right) at
$\sigma=\sigma^*=3.8$. 
(c)~This pattern with $n=7$ is found twice in the data-set. From each 
such subgraph $2$ non-overlapping layered subgraphs 
with $n=4$ and $n=5$ can be generated. 
(d)~The number $p^*(\sigma^*,\mu)$ of subgraphs in the maximum score 
alignment (thick line)
and the fuzziness ${\overline M}^*(\sigma^*,\mu)$ (thin line)
as a function of $\mu$ for $n=5$. 
}
\end{figure}

We now turn to the probabilistic consensus motifs found in the data 
for different number of nodes $n=4$ and $n=5$. 
Fig.~\ref{figmuscan}(a) shows
the $n=4$ consensus motif ${\bar c}_{ij}$ at consecutive values of
 $\mu=\mu^*=3.6$, $\mu=8$, and  $\mu=15$. 
The gray-scale encodes the average number 
of links $\overline{c}_{ij}$ between a given pair of nodes. 
As expected, the fuzziness decreases with increasing values 
of the mismatch penalty $\mu$ and 
$\overline{c}_{ij}$ tends either to zero (no link present) or one 
(link present with certainty) as $\mu \to \infty$. 
The consensus motif is a layered structure, in
this case with $2$ input and $2$ output nodes. 

A similar motif is found for $n=5$.  
Fig.~\ref{figmuscan}(b) shows 
the $n=5$ consensus motif at consecutive values of
$\mu=2.25,5,$ and $12$. As in the 
case of $n=4$, a layered structure is clearly discernible:
the motif consists of $2+3$ nodes forming an input and an
output layer, with links largely going from the input to the output
layer. The left node of the input layer
has an average number of about $30$ outgoing links. These connectivities 
are exceptional since the average out-connectivity of the network is $1.36$. 
 
Comparing the alignments of subgraphs of $n=4$ nodes with those of 
$n=5$ nodes in figure \ref{figmuscan}, one finds 
that many of the subgraphs found in the $n=4$-alignments also are a
part of the subgraphs found in the $n=5$-alignments. This immediately
leads to the question of how to identify {\em larger} patterns in
the network from which the subgraphs at a given value of $n$ are
taken.  Obviously any scoring scheme operating at a fixed number of
nodes $n$ will be blind to the combinatorial possibilities of
selecting subgraphs from a larger pattern. The phenomenon is
exemplified in figure \ref{figmuscan}(c).  
From the $3$-by-$4$ pattern $2$
non-overlapping layered subgraphs with $n=4$ and $n=5$ can be
generated (non-overlapping subgraphs have at most one node in common,
see above). Larger patterns generate correspondingly more
non-overlapping subgraphs. In the supporting text, 
we discuss a simple scheme which allows to identify larger 
patterns as in figure \ref{figmuscan}(c) from smaller subgraphs. 
The pattern of figure \ref{figmuscan}(c) is found twice in the data, 
contributing in total $4$ non-overlapping subgraphs to the alignments with 
$n=4$ and $n=5$.   
The statistics of these patterns at the level of {\em identical 
patterns} has recently been analyzed in \cite{alon_larger}, their 
treatment using {\em probabilistic patterns} remains a future development.

Figure~\ref{figmuscan}(d) shows details of the alignments producing
the consensus motifs at $n=5$, namely, the number of subgraphs
$p^*(\sigma\!=\! \sigma^*, \mu)$ and the fuzziness $\overline
M^*(\sigma\!=\! \sigma^*,\mu)$ plotted as a function of $\mu$.  For
$\mu>12$ the fuzziness reaches zero and the alignment contains $10$
identical, non-overlapping motifs.  This layered pattern has been
found by the approach of ref.~\cite{aloncoli}, which is based on
counting identical motifs.  However, the maximum-likelihood alignment
occurs at $\mu^*=2.25$ and contains a much larger number of $p^* = 24
$ non-overlapping subgraphs, leading to the probabilistic consensus
motif shown on the left in fig.~\ref{figmuscan}(a).  The same effect
is found in the consensus-motif of size $n=4$. Furthermore,
at arbitrary non-zero fuzziness the probability that a given pair of
subgraphs have {\em identical motifs} decreases with subgraph size. As
a result, counting identical motifs, rather than following a
probabilistic approach as the one presented here, will miss an
fraction of relevant subgraphs present in the data which increases with 
the size of the subgraph.

The probabilistic grounding of motif search is also indispensable for
quantitative significance estimates of the results obtained.  Here we
compare the maximum-likelihood alignment in the {\em E.~coli} data set
with suitable random graph ensembles. We do this in two steps, in order
to disentangle the significance of the number of internal links,
and of the mutual similarity of patterns found in the data. 

(i) To  assess the significance of the number of internal links, we 
consider the ensemble of graphs with the same in- and
out-connectivities as the data set but randomly chosen neighbors
\cite{maslovsneppen,berglassig} and compute the distribution of the
score with scoring parameters $\sigma = \sigma^*$, $\mu=0$.
The null distribution of scores from the
randomized graph has the average and standard deviation given by $S^* =
5.7 \pm 2.1$. The score $S^* = 73.1$ found from the data is thus
significantly higher, indicating an enhanced link number with 
respect to the random graph ensemble. (ii) The significance of the
mutual similarity of the aligned patterns is assessed by  comparing the 
data to mutually independent random subgraphs with the same
average density of links. (This null ensemble is generated by
randomizing the internal links of each subgraph independently.) 
We then compute the score with parameters $\sigma = \sigma_0=\sigma^*$ and  
$\mu=\mu^*$ (thereby only focusing on the fuzziness of the
data relative to that found in the ensemble of uncorrelated
subgraphs). This null distribution of scores has average and standard
deviation given by $S^* = 27.1 \pm 6.3$; the corresponding score $S^* = 50.1$ 
found from the
data is thus significantly higher. We note that the assessment 
of subgraph similarity is quite subtle. Subgraphs taken from a large 
but finite random graph may show a `spurious' mutual similarity with 
respect to independent random subgraphs due to a prevalence of internal loops.
 
The statistical significance of the results can be formulated more
precisely using so-called $p$-values, which involve the tail of the
score distribution in the random graph ensemble.  Fast and reliable
$p$-value estimates are crucial for the search in large databases, as
it is well known for sequence alignment~\cite{karlinaltschul}. This
approach can be carried over to the graph alignments discussed
here. 

The statistical framework presented is very flexible.  For example, as
large-scale data on the logic of gene regulation become available, the
definition of the pairwise mismatch (\ref{pairwise_mismatch}) can be
extended to reward aligning sets of nodes performing the same logical
function. In this way, features of motifs going beyond their topology
can be explored.  Similarly, simple modifications of the mismatch
score allow the analysis of undirected networks, networks whose links
have a specific function (repressive or enhancing) or whose
interaction strength is quantified by a real number. The statistical
framework presented is very flexible.  For example, as large-scale
data on the logic of gene regulation become available, the definition
of the pairwise mismatch (\ref{pairwise_mismatch}) can be extended to
reward aligning sets of nodes performing the same logical function. In
this way, features of motifs going beyond their topology can be
explored.  Similarly, simple modifications of the mismatch score allow
the analysis of undirected networks, networks whose links have a
specific function (repressive or enhancing) or whose interaction
strength is quantified by a real number.

The prospect of a sizable amount of new data on biological networks
becoming available over the next few years through high-throughput
methods opens exciting opportunities to identify the building blocks
of molecular information processing in a wide range of organisms,
and even build phylogenetic histories of regulation
from transcription network data.

\vspace{0.5cm}
\noindent {\bf Acknowledgments:}
Many thanks to T. Hwa for fruitful discussions and to U. Alon 
and P. Arndt for comments on the manuscript. 
This work has been supported through DFG grant LA 1337/1-1. 
We thank the KITP Santa Barbara, ICTP Trieste, the Centro di Ricerca
Matematica Ennio De Giorgi, Pisa, and the Aspen Center for Physics
for hospitality during various stages of this work.


\protect{\newpage}
\pagestyle{empty}
\subsubsection*{Supporting Text}

In the supplementary material we first give additional details on the
alignment algorithm and then discuss a simple procedure for
identifying larger modules generating multiple smaller subgraphs.
  
The algorithm proceeds in four stages:
\begin{enumerate}
\item By enumeration all unique non-treelike subgraphs of size $n$
  are found. We consider only subgraphs where each node carries at
  least two internal links, other than self-links (``exclusion of
  dangling links'').  The reason is that including dangling links
  would generate from each subgraph an artificially inflated family of
  subgraphs generated by including all combinations of neighboring
  nodes into the subgraph.  The enumeration is done by first finding
  all closed paths in the graph of length shorter than $2n-3$.  (The
  maximum length derives from considering the non-treelike structure
  with the longest pathlength from the origin through all points of
  the subgraph back to the origin.  The graph is considered as
  undirected at this stage.)  The subgraphs are labeled by $\alpha =
  1, \dots, p^{\max}$.

\item The pairwise minimal mismatch $M^{\alpha \beta}$ for all
  \emph{pairs of subgraphs} $\alpha,\beta$ is found by enumerating all
  $n!$ possible alignments of each pair of subgraphs.  For each pair
  of subgraphs $\alpha$ and $\beta$ we determine whether they overlap
  by counting the number of nodes they have in common.  The elements
  of the coupling matrix $\tilde{M}^{\alpha \beta}$ in the Hamiltonian
  {\bf \ref{hamiltonian}} are given by $M^{\alpha \beta}$ if the subgraphs
  do not overlap, and by a large number, chosen to be $10$, if they
  do.

\item The next task is to select a subset of the subgraphs such that
  the total score {\bf \ref{align_score}} is maximized at given values of
  the scoring parameters.  To this end simulated annealing is used,
  with the (negative) score as the energy function, increasing the
  inverse temperature from $0$ to $10$ in $1000$ Monte-Carlo sweeps.
  We assign each {\it subgraph} a spin variable: spin $s^{\alpha}=1$
  implies that the subgraph $\alpha$ is included in the alignment,
  spin $s^{\alpha}=0$ that it is not.  The contribution to Eq. 
  {\bf \ref{align_score}} from the mismatch of two subgraphs 
  acts as a coupling between their spins, the
  contribution of subgraph $\alpha$ to the total number of links $L$
  in {\bf \ref{align_score}} acts as a local field, resulting in the
  Hamiltonian {\bf \ref{hamiltonian}}.  The evaluation of the last term in
  {\bf \ref{align_score}}, $\log (Z/Z_0)$, is discussed below.
\end{enumerate}

The last step is repeated at different values of $\sigma$ and $\mu$ in
order to perform the parametric optimization leading
to Fig. \ref{figopt}{\it b}. The parameter $\sigma_0$ describing the
null ensemble, on the other hand, is determined independently by
considering the non-treelike subgraphs found in the randomized graph as
described in the paper. $\sigma_0$ is chosen such that the average number
of internal links in the ensemble of uncorrelated subgraphs with an
enhanced number of links {\bf \ref{sigma_enhanced}} equals that of 
the non-treelike subgraphs found in the randomized network, 
\begin{equation*}
\frac{1}{p} \sum_{\alpha = 1}^p 
\langle L ({\bf c}^\alpha) \rangle_{\sigma_0, \mu=0} = \overline L_{\text{randomized}}  \ .
\end{equation*}
Note that the ensemble {\bf \ref{sigma_enhanced}} still depends 
on the connectivities of the nodes in each subgraph. 
The generalization to 
several \emph{groups of subgraphs}, where only subgraphs from the 
same group are aligned with each other, 
can be done by admitting more states of the Potts-like spins
$s^{\alpha}$. For $q$-state spins this would group the subgraphs
into $q-1$ clusters much like in super-paramagnetic
clustering~(1).

There are two approximations behind this algorithm.
First, treelike subgraphs are excluded from the start. This step cuts down
an enormous number of combinatorial possibilities associated with treelike
subgraphs, which, different connectivities apart, are always locally
similar.

Second, it uses the minimal mismatch obtained from the {\it pairwise}
alignment of subgraphs (step $2$), even though the minimal mismatch
obtained by aligning a set of more than two subgraphs may be higher
than that of the sum of pairwise alignments. This is easily seen by
comparing the total number of alignments of all \emph{pairs} chosen
from $p$ subgraphs, $(n!)^{p(p-1)/2}$, with the number of alignments
of $p$ subgraphs, $(n!)^p$. However, in the case of interest, where
the graph contains multiple copies of a motif (possibly corrupted by
noise), the sum of pairwise minimal mismatches will typically be very
close to the minimal mismatch obtained from aligning all subgraphs
simultaneously.

The maximal-score alignment
$\A^{\star}(\sigma, \mu)$ turns out to be unique in most subgraphs.  To see
this, consider all alignments $\A^\alpha$ of a particular subgraph
$\alpha$ with respect to the other subgraphs whose alignment is kept
fixed. Two different alignments have the same score if and only if
there is a permutation of the nodes ${r_1^\alpha, \dots, r_n^\alpha}$
leaving both the adjacency matrix $c^\alpha_{ij}$ and the matrix
$w^\alpha_{ij}$ defined above Eq.~{\bf \ref{single_null}} invariant.
While symmetries of the adjacency matrices occur frequently, entries
of the matrix $w_{ij}$ are unique in most subgraphs, since the
connectivities in biological networks are broadly distributed. 

We now discuss the normalizing constant {\bf \ref{Z_musigma}}
of the alignment ensemble {\bf \ref{q_alignment}}. We approximate 
the likelihood of given parameter values, which involves the sum over 
all alignments $\A$ as 
in {\bf \ref{Z_musigma}}, by the corresponding maximum-score alignment. (In the
literature for sequence alignment, this is known as the Viterbi approximation.)
An improved likelihood estimate is possible using probabilistic graph 
alignment algorithms but is not expected 
to alter our results qualitatively. 
The optimal alignment
has $p^{\star} \equiv p^{\star}(\sigma^{\star},\mu^{\star})$ subgraphs 
with average 
internal link number 
$\overline L^{\star} \equiv \overline L^{\star} (\sigma^{\star}, \mu^{\star})$ 
and
fuzziness 
$\overline M^{\star} \equiv  \overline M^{\star} (\sigma^{\star}, \mu^{\star})$. 
As may be seen by differentiation of {\bf \ref{align_score}} with respect 
to the scoring parameters, at $\sigma=\sigma^{\star}$ and $\mu=\mu^{\star}$ 
the $Q$ ensemble fits to the data set in the sense that the expectation
values of the  internal link number  and the fuzziness equal the actual
values,
\begin{eqnarray*}
\label{eq_linkno_fuzz}
\frac{1}{p} \sum_{\alpha = 1}^p 
\langle L ({\bf c}^\alpha) \rangle_{\sigma^{\star}, \mu^{\star}} = \overline L^{\star},
\\
\frac{1}{p^2} \sum_{\alpha,\beta = 1}^p
      \langle M({\bf c}^\alpha, {\bf c}^\beta) \rangle_{\sigma^{\star}, \mu^{\star}} 
= \overline M^{\star}.
\end{eqnarray*}

The normalizing constant {\bf \ref{Z_musigma}} needs to be 
computed for two sets of parameters; for $\sigma$, $\mu$ characterizing 
the $Q$ ensemble, and for $\sigma_0$, $\mu_0$ characterizing the 
$Q_0$ ensemble. In both cases the normalizing constant consists 
of a trace over
the link configuration $\{ {\bf c}^{\alpha} \}$ in all subgraphs.
Since the constant {\bf \ref{Z_musigma}}
factorizes in the link labels $i,j$, we
consider only a single of these factors (a single ``string''), drop the
$i,j$ indices, and separate the bilinear form of the pairwise
mismatch {\bf \ref{pairwise_mismatch}} into a quadratic and a linear part. 

Formally, this expression is the partition function of a mean-field
ferromagnet in a fluctuating field. The field depends on the local
connectivities of each node along the ``string'' via the ensemble
$P_0$, Eq. {\bf \ref{single_null}}.
Using a Hubbard-Stratonovich transformation to
linearize the quadratic term, the trace over $\{ c^{\alpha} \}$ can
be performed, giving
\begin{equation*}
\label{spe}
Z = \int \frac{dt}{\sqrt{2 \pi/p}}
\exp \left\{ -p t^2/2 + 
\sum_{\alpha}^p g^{\alpha}(t) \right\} \ , {\bf [12]}
\end{equation*}
where 
\begin{equation*}
g^{\alpha}(t)=\log \left[
(1-w^{\alpha} ) + w^{\alpha} \exp \left\{ \sqrt{2 \mu}t+\sigma-\mu  \right\}
\right] \ .
\end{equation*}
For large $p$ this expression can be evaluated by a saddle point integral,
giving
\begin{equation*}
\label{saddle_point}
\log Z \approx -p {t^{\star}}^2/2 + \\
\sum_{\alpha}^p g^{\alpha}(t^{\star}) + {\cal O}(\log p) \ ,
\end{equation*}
where $t^{\star}$ maximizes the exponent in Eq.{\bf 12}. The contribution to
leading order of adding a new subgraph with index $\alpha$ is thus
\begin{equation*}
\label{delta_F}
\Delta \log Z \approx
-{t^{\star}}^2/2 + g^{\alpha}(t^{\star}) \ .
\end{equation*}
The change of $t^{\star}$ as a finite number of subgraphs is added to or
removed from the alignment alters the result {\bf \ref{delta_F}} only by
terms of order $p^{-1}$. It thus turns out to be sufficient to update the
saddle-point value $t^{\star}$ for each link $i,j$ once per Monte-Carlo
sweep of the algorithm.

In order to compute {\bf \ref{saddle_point}} for each pair $i,j$ in the
Viterbi approximation, the one-to-one mapping between nodes in each
subgraph $\A$ is needed, going beyond the {\em pairwise} alignment.
This mapping is also needed to produce the plots of the consensus
motifs in Fig. \ref{figmuscan}.  It is produced by minimizing the
fuzziness over the mapping between nodes in each subgraph, again using
Monte-Carlo dynamics ($100$ Monte-Carlo sweeps while linearly
increasing the inverse temperature from $0$ to $10$). The result of
course depends on the subgraphs in the alignment, and thus the mapping
ought to be updated each time a subgraph is added or removed from the
alignment. In practice, however, one update of the mapping between
nodes in each subgraph every $250$ steps of the algorithm is
sufficient. The reason for this is again that the mapping between
nodes in subgraphs in the alignment is unchanged as motifs
sufficiently close to the consensus motif enter/leave the alignment.

Finally, we discuss a simple procedure for
identifying these structures {\em from the set of subgraphs at fixed
  (small) $n$}. First, for a given subgraph of size $n$, all neighbors
with at least two links to the subgraph are enumerated. In this way,
non-treelike subgraphs without dangling bonds with $n+1$ nodes are
generated.  This procedure is repeated for the entire list of $p$
subgraphs. Several subgraphs of size $n+1$ will occur repeatedly in this list:
the more subgraphs of size $n$ can be be generated from the larger
$n+1$ subgraph the more frequently it occurs on the list. Thus ranking the
$n+1$ subgraphs according to the number of times they occur in the
list, one obtains the $n+1$ subgraph from which the largest number of
$n$ subgraphs derive. Clearly this procedure can be repeated
iteratively, leading to subgraphs of increasing size.  
In fact, Fig. \ref{figmuscan} {\it c} is the result of applying this
scheme to the non-treelike subgraphs with $n=5$. Iterating twice, one
finds two instances of the $n=7$ layered structure of Fig.
\ref{figmuscan}{\it c}. 

\vspace{.5cm}
\noindent
[1] Blat,M., Wiseman,S., \& Domany,E. (1996)
{\it Phys. Rev. Lett.} {\bf 76} 3251-3255.

\begin{thebibliography}{100}

\bibitem{durbin}Durbin,R., Eddy,S.R., Krogh,A., \& Mitchison,G. (1998)
{\it Biological sequence analysis}, (CUP, Cambridge, UK).

\bibitem{davidson}Davidson,E.H. (2001)
{\it Genomic regulatory systems: Development and Evolution}
(Academic Press).

\bibitem{tautz.review}Tautz, D. (2000)
{\it Current Opinion in Genetics \& Development}
{\bf 1}, 575-579.

\bibitem{uetz}Uetz,P., Giot,L., Cagney,G., Mansfield,T. A.,
Judson,R. S., Knight J.R., Lockshon D., Narayan V., Srinivasan M., 
Pochart P., Qureshi-Emili A., Li Y. et al.(2000)
{\it Nature} {\bf 403}, 623-627.

\bibitem{lockhart}Lockhart,D.J., \& Winzeler,E.A. (2000)
{\it Nature} {\bf 405},827-836.

\bibitem{aloncoli}Shen -Orr,S., Milo,R., Mangan,S., \& Alon,U. (2002)
{\it Nature Genetics} {\bf 31} 64-68.

\bibitem{alonyeast}
Milo,R., Shen-Orr,S., Itzkovitz,S., Kashtan,N., Chklovskii,D. \& Alon,U. (2002)
{\it Science}, {\bf 298} 824-827.

\bibitem{newman_p}
Newman,M.E.J., Strogatz,S.H., \& Watts,D.J. (2001)
{\it Phys. Rev.} {\bf E 64}, 026118.

\bibitem{newman_q}
Newman,M. (2002)
{\it Phys. Rev. Lett.} {\bf 89}, 208701.

\bibitem{berglassig}
Berg,J., \& L\"assig,M. (2002)
{\it Phys. Rev. Lett.} {\bf 89}, 228701.


\bibitem{molclock}
Costas,J., Casares,F., \& Viera,J. (2003)
{\it Gene} {\bf 310}  215-220.

\bibitem{stormo}Stormo,G.D., \& Hartzell,G.W. (1989)
{\it Proc. Nat. Acad. Sci. USA} {\bf 86}, 1183-1187.

\bibitem{bussemaker}Bussemaker,H.J., Li,H., \& Siggia,E.D. (2000)
{\it Proc. Nat. Acad. Sci. USA} {\bf 97}, 10096-10100.

\bibitem{felsenstein}
Felsenstein,J. (1981)
{\it Journal of Molecular Evolution} {\bf 17} 368-376. 

\bibitem{ullmann}
Ullmann,J.R. (1976)
{\it Journal of the Association for Computing Machinery} {\bf 23}
31-42.

\bibitem{messmerbunke}
Messmer,B.T.,\& Bunke,H. (1998)
{\it IEEE Trans. on Pattern Analysis and Machine Intelligence}
{\bf 20 (5)} 493-504.

\bibitem{gareyjohnson}Garey,M.R. \& Johnson,D.S. (1979)
{\it Computers and Intractability: A Guide to the Theory of Np-Completeness}
(Freeman,NY).

\bibitem{maslovsneppen}
Maslov,S., \& Sneppen,K.
(2002) {\it Science} {\bf 296} 910-913.

\bibitem{alonsubgraph}
Itzkovitz,S., Milo,R., Kashtan,N., Ziv,G., \& Alon,U. (2003)
{Phys. Rev.}{\bf E 68}, 026127.

\bibitem{radicetal}
Berg, J., L\"assig,M., \& Radic,S. (2003)
http://arxiv.org/abs/cond-mat/0301574 .

\bibitem{karlinaltschul}
Karlin,S., \& Altschul,S.F (1990)
{\it Proc. Nat. Acad. Sci. USA} {\bf 87} 2264-2268.

\bibitem{domany}Blat,M., Wiseman,S., \& Domany,E. (1996)
{\it Phys. Rev. Letters} {\bf 76} 3251-3255.

\bibitem{alon_larger}
Kashtan,N., Itzkovitz,S., Milo,R., \& Alon,U. (2003), 
http://arXiv.org/abs/q-bio/0312019 

\end{thebibliography}
\end{document}